\documentclass[twocolumn,showpacs,preprintnumbers,amsmath,amssymb,prb,nofootinbib]{revtex4}
\usepackage{amsfonts,amssymb,latexsym,amsmath,amscd}
\usepackage{times,mathptm,color}
\usepackage{anysize}
\usepackage{graphicx}
\usepackage[ps2pdf,colorlinks]{hyperref}

\newcommand{\ket}[1]{|#1\rangle}
\newcommand{\bra}[1]{\langle#1|}

    {
    \smallskip
    \refstepcounter{theorem}
    \noindent
    {\bf Example \arabic{section}.\arabic{theorem}} \ \ }
    {\hspace*{\fill}{$\Diamond$}
    \smallskip}

    {
    \smallskip
    \refstepcounter{theorem}
    \noindent
    {\bf Remark \arabic{section}.\arabic{theorem}} \ \ }
    {\hspace*{\fill}{$\Diamond$}
    \smallskip}

    {
    \smallskip
    \refstepcounter{theorem}
    \noindent
    {\bf Definition \arabic{section}.\arabic{theorem}} \ \ }
    {\hspace*{\fill}{\ }
    \smallskip}

    {
    \noindent
    {\bf Proof{#1}:  }
    }
    {\hspace*{\fill}{$\Box$}\smallskip}

    {
    \noindent
    {\bf Sketch{#1}:  }
    }
    {\hspace*{\fill}{$\Box$}\smallskip}

    {
    \noindent
    {\bf Reference:  }
    }
    {\hspace*{\fill}{$\odot$}\smallskip}

\begin{document}
\bibliographystyle{apsrev}


\title{Single qubit gates with a charged quantum dot using minimal resources}

\author{Fran\c{c}ois Dubin$^{\ast}$ and Gavin Brennen$^{\dagger}$}
\affiliation{$^{\ast}$Institute for Experimental Physics, University of
Innsbruck,
Technikerstr.\ 25, A-6020 Innsbruck, Austria \\
$^{\dagger}$Institute for Quantum Optics and Quantum Information of the
Austrian Academy of Sciences, 6020 Innsbruck, Austria}

\date{\today }
\pacs{78.67.Hc, 03.67.Lx}
\begin{abstract}
We investigate coherent control of a single electron trapped in a semiconductor
quantum dot.  Control is enabled with a strong laser field detuned
with respect to the electron light-hole optical transitions. For a
realistic experimental situation, i.e. with a weak magnetic field
applied along the growth direction, high fidelity arbitrary
rotations of the electron spin are possible using a single laser
spatial mode.  This makes viabile quantum gates with 
electron spins in systems with restricted optical resources.
\end{abstract}

\maketitle

\section{Introduction}
A large research effort is presently devoted to the
manipulation of an increasing number of quantum systems in order
to use them as a possible support for quantum computation. In
atomic physics, on the one side single trapped ions have lead to
major steps toward large scale quantum hardware. The almost
perfect control of trapped ions motional and electronic long lived
states has enabled high fidelity single qubit operations
\cite{Gulde}, very efficient two qubit gates \cite{Ferdi} and
recently the realization of the first quantum-bite \cite{Hartmut}.
On the other side, cold atomic ensembles are also very attractive
and have emerged as an alternative light-matter interface.
Quantum state transfer from an atomic cloud onto a photonic qubit
has been demonstrated \cite{Matsukevich1} as well as the
entanglement of two remote atomic ensembles \cite{Matsukevich2}.

In solid state physics, charged semiconductor quantum dots are
also promising candidates to implement quantum computing
protocols. A single trapped electron in a quantum dot has
long lived spin states \cite{Pal} which in principle allow for
long information storage times. Very recently, the spin state of a
single electron trapped in a self assembled quantum dot has been
prepared with a fidelity close to unity \cite{Attature}. This
initialization, prerequisite to any spin manipulation, is a major
advance which can be seen as a laser-cooling mechanism implemented
in a solid-state environment. In the experiments of Attat\"{u}re
and coworkers, a weak magnetic field ($\approx$ 100 mT) was
applied along the quantum dot growth direction (Faraday
configuration) such that the electron spin state becomes less
sensitive to surrounding fluctuations. In many other works, single
charged quantum dots are also studied when a strong magnetic field
($\approx$ 5-10 T) is applied perpendicular to the growth axis
\cite{Shabaev} (Voigt configuration). The latter removes the degeneracy of the two spin projections
along the field which are then the basis for subsequent manipulations. In the Voigt configuration electron
spin coherences have been measured \cite{Sham1} while
theoretical proposals have shown how optical control can be
achieved \cite{Sham2,Sham3}.

A trapped electron spin can be rotated in different ways, in general due to
subtle fermion exchanges which can take place with virtual
excitons coupled to unabsorbed photons \cite{Monique1}.  In quantum dots, when
a strong magnetic field is applied in the Voigt configuration,
Raman transitions have been proposed to perform arbitrary single
qubit operations \cite{Sham2,Sham3}. Using two crossed polarized
and detuned laser pulses, the electron spin states can be
virtually coupled to ''trion" states\cite{note}. The trapped
electron spin can therefore be well manipulated, with a close to
unity fidelity for typical operation times of 50 ps. A
procedure to perform general spin rotations via Stimulated Raman
Adiabatic Passage (STIRAP) was also proposed \cite{Elisa}, with a
required auxiliary ground state, to perform single qubit
operations and quantum gates in coupled quantum dots.

In this work, we show that the spin state of a single charged
quantum dot can also be completely controlled with a single laser
frequency, with or without a weak magnetic field applied in the
Faraday configuration. The latter situation corresponds to
experiments which have up to now provided remarkable results,
furthermore, removing the degeneracy protects the qubit from spin flipping errors.
 Using a single laser beam, of prepared elliptical polarization and tilted with
respect to the quantum dot growth direction, a coherent coupling
between the two electron spin states is created. Arbitrary
rotations can thereby be obtained in few tens of picoseconds, with
high fidelities for realistic experimental parameters. We derive
the photon induced entanglement of the electron spin states
through an effective Hamiltonian describing the virtual
transitions towards trion states.  The mechanism is highlighted by
the recently developed Shiva diagrams for interacting composite
bosons \cite{Monique1}.

Our proposal is based on the semiconductor light-hole excitations.
In a GaAs quantum well, the latter have allowed the experimental
demonstration of an induced electron spin coherence, in a
waveguide geometry i.e. with an excitation laser propagating in
the plane of the quantum well \cite{Sarkar}. In the experiments of
Sarkar et al., a light-hole relaxation time of about 40 ps is
deduced. This value at first seems to prohibit the use of
light-holes for any electron spin manipulation. The same order of magnitude is 
indeed expected for self-assembled quantum dots  (to our best
knowledge, in quantum dots no experimental values for the
light-hole lifetime have been reported). However, we will demonstrate in the following
that high-fidelity qubit gates can be achieved using light-holes. In the worst case scenario 
for which light-holes would
have a lifetime of 50 ps in a quantum dot, we show that  arbitrary
rotations of the electron spin already have at least 97$\%$
fidelities with a gate time of few tens of picoseconds.

In strained
quantum dots, light-hole levels can become the lowest energy hole
states \cite{lh}. This suppresses most relaxation channels and
yields long lived light-holes. Assuming such a quantum dot, Calarco
and co-workers \cite{PZ} have proposed to use Raman transitions
between electron and light-hole states to perform single qubit rotations
and two qubit gates. With two orthogonal laser fields, of linear
and circular polarizations, the electron spin states can indeed be
coherently coupled. In this work, we show that the resulting
complex two orthogonal axes geometry is in fact not required, a single laser
beam of suitable elliptical polarization and propagation direction is sufficient.
Using a single beam provides more optical access for scattered light
which improves photon detection efficiency.

\section{Setup}

\begin{figure}
\includegraphics[width=6cm]{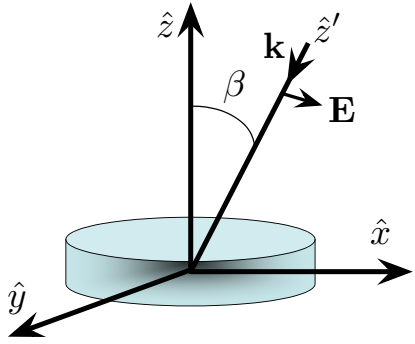}
\caption{Sketch of the considered experimental configuration.
The laser beam propagates in the $(Oxz)$ plane of a
self-assembled quantum dot, at an angle $\beta$ with respect to
the $\hat{z}$ axis. The laser polarization ${\bf E}$ is in general elliptical and
has projection along
$\hat{x},\hat{y},\hat{z}$.}
\label{fig1}
\end{figure}

\begin{figure}[!]
\centerline{\scalebox{0.4}{\includegraphics{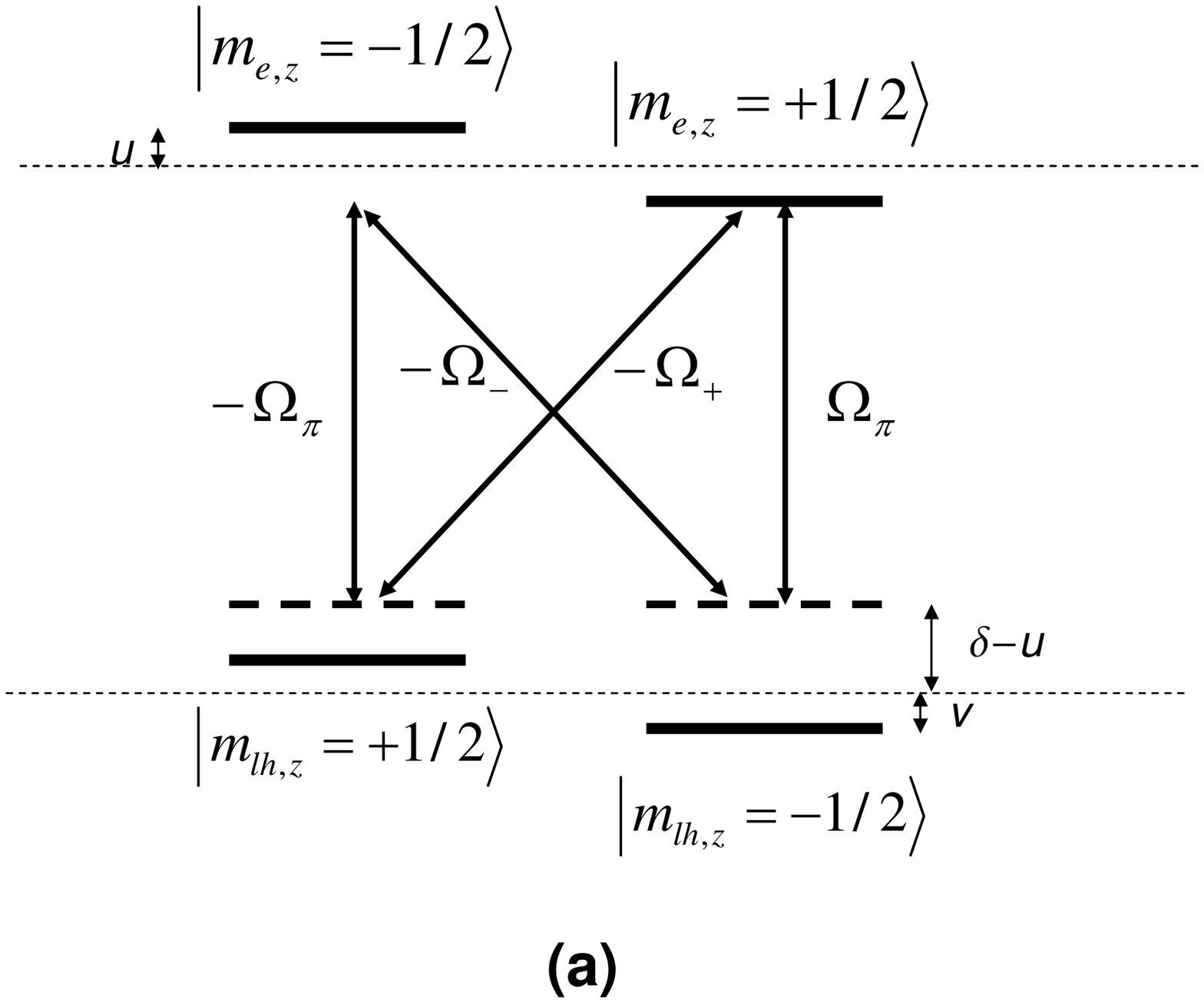}}}
\vspace*{.5cm}\centerline{\scalebox{0.4}{\includegraphics{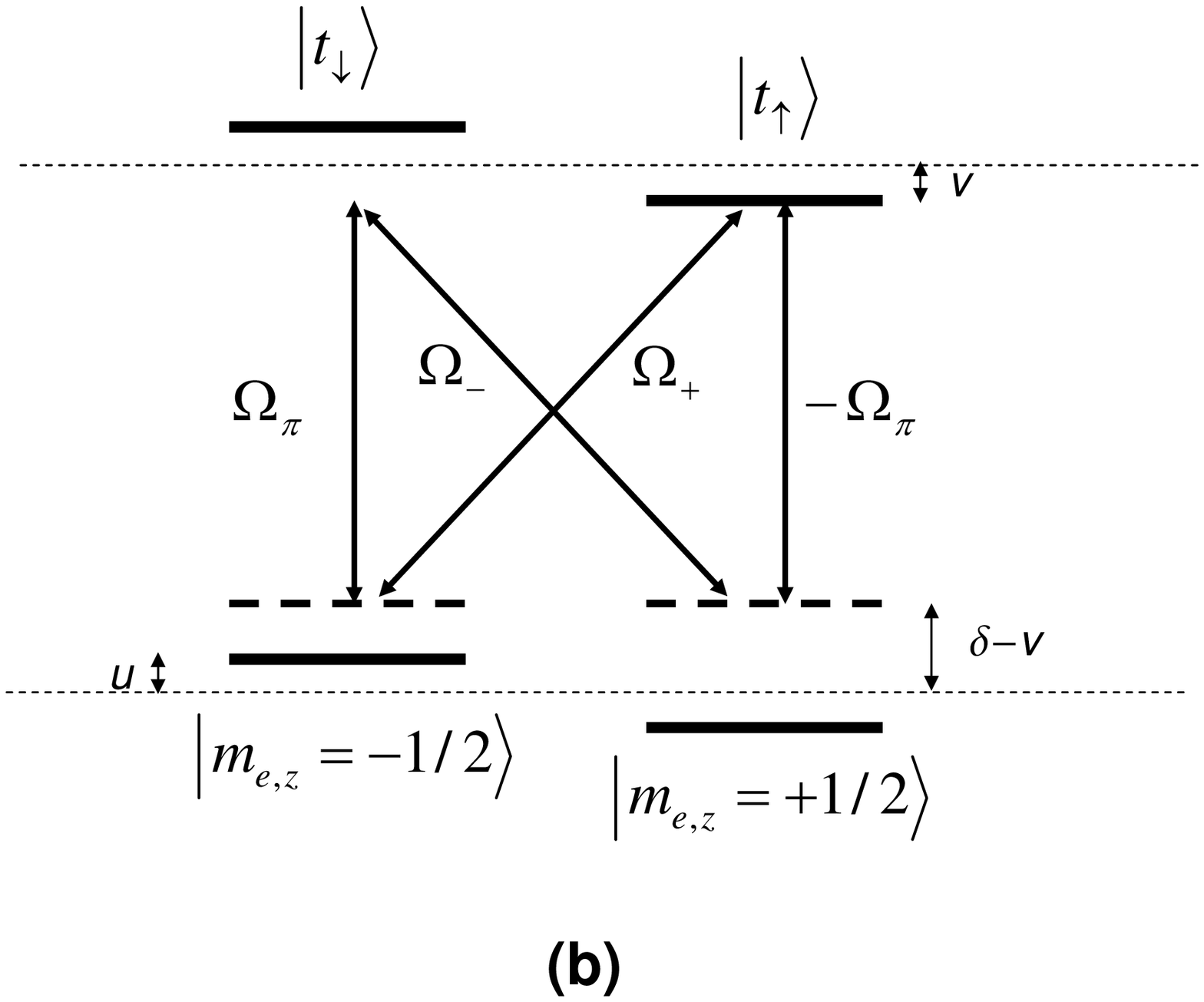}}
} \caption{(a) Optically allowed transitions between the electron
and the light-hole levels. The Rabi frequencies are defined in the
text. (b) Optical transitions when considering that the electronic
states are coupled to trions. Note that in this second
representation the different coupling strengths have different
signs.}\label{fig2}
\end{figure}

As shown in Figure \ref{fig1}, we consider an excitation laser
which propagates in the (Oxz) plane of a self-assembled quantum
dot at an angle $\beta$ relative to the $\hat{z}$ axis.
 In the most general situation, the laser field
polarization has non vanishing projections along all directions,
$\hat{x}$, $\hat{y}$, and $\hat{z}$. The first two are used to
couple light-hole levels to the two electronic levels via
$\sigma^+$ and $\sigma^-$ optical transitions. The last projection
also couples the light-hole levels to the electronic ones but via
so called $\pi$ transitions maintaining  the projection of the
angular momentum along (Oz) (see Figure \ref{fig2}.a).
Consequently, a coherent coupling between the trapped electron
states, with up and down spin projections along (Oz), can be
induced. Moreover, since semiconductors have a refraction index
of the order of 3.5, the angle can be $|\beta|\lesssim
20^{\circ}$.

\begin{figure}
\begin{center}
\includegraphics[width=\columnwidth]{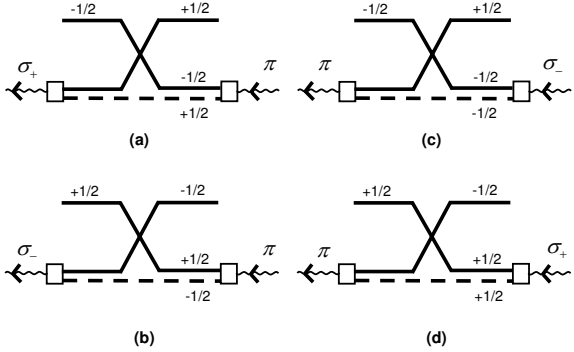}
\end{center}
\caption{Diagrammatic representation of the physical processes
involved in the electron spin-flip. The wavy line represents pump
photons, the full and dashed lines correspond to the electron and
light-hole respectively. (a) Exchange interaction leading to the trapped
electron spin-flip and the transformation of a $\pi$ photon into a
$\sigma^+$ one. (b) $\sigma^-$ photons can also be obtained from
$\pi$ ones while flipping the electron spin. (c) and (d):
Electron spin-flip obtained from the transformation of $\sigma$
photons into $\pi$ ones.}\label{fig3}
\end{figure}

The carrier exchange between the trapped real electron and the
virtual light-hole-electron pair, coupled to unabsorbed $\pi$,
$\sigma_+$ or $\sigma_-$ photons, allows to switch the trapped
electron spin projection along (Oz) (see Figure \ref{fig3}). This
switch imposes the transformation of a $\pi$ photon into a
$\sigma_+$ or a $\sigma_-$ one, or vice-versa, in order to
conserve the total momentum of the system. Note that, in these
processes, the intermediate states are made of one real and one
virtual electron with opposite spins, along with one virtual
light-hole. The optical spin manipulation we propose then implies
the formation of a semi virtual trion, in the same way as in Ref
\cite{Sham3}.

\section{Formalism for the spin flip of a trapped electron}

\subsection{Photon and dot Hamiltonians}

We consider unabsorbed photons with a detuning $\delta$ with
respect to the electron-light-hole transition, i.e. with a
frequency
\begin{equation}
\omega=E_e+E_{lh}+\delta, \label{eq1}
\end{equation}
$E_e$ and $E_{lh}$ being the electron and light-hole energies
without applied magnetic field respectively. In self assembled
quantum dots, the energy splitting between the heavy and the
light-holes levels is typically of the order of tens of meV,
while the energy splitting between the confined electronic levels
can be made much larger for small dot structures \cite{Cusack}.
Consequently, in the following we only consider couplings between
the light-holes and the first confined electronic state,  the
laser photons being assumed far detuned with other electronic transitions.

In a quantum dot small compared to the exciton Bohr radius, the
Coulomb interaction is rather unimportant, the carrier energy
being controlled by localization. In addition, as already shown on
various examples\cite{Monique2}, the optical non linearities are
driven by pure carrier exchanges, due to a bare dimensionality
argument, so that Coulomb interaction between carriers is going to
play a minor role in the problem investigated here. This is why
Coulomb interactions can be neglected in the dot Hamiltonian
reducing to
\begin{equation}
H_d=\sum_{s=\pm1/2}\hbar\omega_s
a_s^{\dag}a_s+\sum_{s=\pm1/2}\hbar\omega_{\frac{3}{2},s}b_{\frac{3}{2},s}^{\dag}b_{\frac{3}{2},s}.
\end{equation}\label{eq2}
$a$ ($a^{\dag}$) and $b$ ($b^{\dag}$) being fermionic operators
for the destruction (creation) of a trapped electron or hole
respectively.

\subsection{Dot-photon interaction}

The semiconductor-laser-photons interaction is in general written
as $W$=$V$+$V^{\dag}$, where
$V^{\dag}$=$V_+^{\dag}$+$V_-^{\dag}$+$V_0^{\dag}$ with
\begin{equation}
\begin{array}{lll}
V_{+}^{\dag}&=&\lambda_{+}\left(a_{-
\frac{1}{2}}^{\dag}b_{\frac{3}{2},\frac{3}{2}}^{\dag}-
\sqrt{\frac{1}{3}}a_{
\frac{1}{2}}^{\dag}b_{\frac{3}{2},\frac{1}{2}}^{\dag}-
\sqrt{\frac{2}{3}}a_{\frac{1}{2}}^{\dag}b_{\frac{1}{2},\frac{1}{2}}^{\dag}\right)\\
V_{-}^{\dag}&=&-\lambda_{-}\Big(a_{\frac{1}{2}}^{\dag}b_{\frac{3}{2},-\frac{3}{2}}^{\dag}+
\sqrt{\frac{1}{3}}a_{-
\frac{1}{2}}^{\dag}b_{\frac{3}{2},-\frac{1}{2}}^{\dag}\\
&&-
\sqrt{\frac{2}{3}}a_{- \frac{1}{2}}^{\dag}b_{\frac{1}{2},-\frac{1}{2}}^{\dag}\Big)\\
V_0^{\dag}&=&\sqrt{\frac{2}{3}}\lambda_0\Big(a_{\frac{1}{2}}^{\dag}b_{\frac{3}{2},-\frac{1}{2}}^{\dag}
-a_{-\frac{1}{2}}^{\dag}b_{\frac{3}{2},\frac{1}{2}}^{\dag}+a_{\frac{1}{2}}^{\dag}b_{\frac{1}{2},-\frac{1}{2}}^{\dag}\\
&&
+a_{-\frac{1}{2}}^{\dag}b_{\frac{1}{2},\frac{1}{2}}^{\dag}\Big)
\end{array}
\end{equation}\label{eq3}
The photon field is treated classically, with the definition
$\lambda_{0,\pm}=ePA_{0,\pm}/\omega mc$
\cite{Monique3}, where ${\bf A}$ is the field vector potential, $m$ is the electron mass, and
$P$ is the canonical momentum whose vector components are
identical.  Restricting to the light-hole levels, we only
consider couplings employing the $b_{\frac{3}{2},\pm\frac{1}{2}}$
operators.

The coupling $W$ does not conserve the number of carriers, the
changes it induces to the degenerate zero, $|vac\rangle$, and one
electron states $|\pm 1/2\rangle$ only appear at second order
in this coupling. In order to derive the dynamics in the
$m_s=\pm 1/2$ trapped electron states in the presence of pump
photons, we express an effective Hamiltonian, $H_{\rm eff}$, to obtain
proper equations of motion. In this approach, several constraints
are imposed. Apart from hermiticity, most importantly, $H_{\rm eff}$
must have the same eigenvalues as the original Hamiltonian,
$H$=$H_d$+$W$, with the same degeneracy. $H_{\rm eff}$ is in fact
obtained from an appropriate unitary transformation \cite{Cohen},
and reads at second order in the coupling $W$
\begin{equation}
\begin{array}{lll}
H_{\rm eff}&=& H_d +\frac{1}{2}\displaystyle{\sum_k\sum_{s,s'}} \ket{s}\bra{s}W|k\rangle\langle
k|W\\
&&\Big( \frac{1}{E_s+\omega-E_k}
+\frac{1}{E_{s'}+\omega-E_k}\Big)\ket{s'}\bra{s'}
\label{eq4}
\end{array}
\end{equation}
where the states $|k\rangle$ correspond to any intermediate states
and $\ket{s},\ket{s'}$ are initial and final ground states
respectively.

\subsection{Couplings of the zero electron states}

\begin{figure}
\includegraphics[width=9cm]{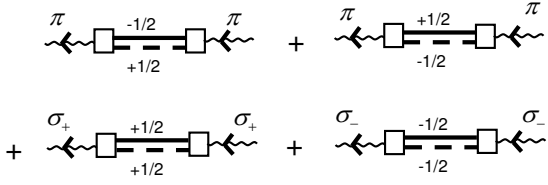}
\caption{Different contributions to the energy change of the zero
electron state, $|vac\rangle$.} \label{fig4}
\end{figure}

The dot-photon interaction induces an energy shift of the
zero-electron state, $|vac\rangle$. It is given by the matrix
elements of $H_{\rm eff}$ in this degenerate subspace. From eqs.
(\ref{eq1}-\ref{eq4}) it is easy to show that this operator is
diagonal, with
\begin{eqnarray}
\langle vac|H_{\rm eff}|vac\rangle=
\frac{2}{3}\frac{|\lambda_0|^2}{\delta+(u+v)}+\frac{2}{3}\frac{|\lambda_0|^2}{\delta-(u+v)}\nonumber\\
+\frac{1}{3}\frac{|\lambda_+|^2}{\delta+(u-v)}
+\frac{1}{3}\frac{|\lambda_-|^2}{\delta-(u-v)}\label{eq5}
\end{eqnarray}
$u$ and $v$ being the electron and light-hole energy shifts
induced by the applied magnetic field. Note that these are very
small compared to $\delta$ for our parameter regime of interest. 
We remark that two
channels exist for the coupling of $\pi$ photons with the dot
electron-light-hole states, while there is only one channel for
the circularly polarized photons. The different terms of
eq.(\ref{eq5}) are shown in Figure \ref{fig4}.

\subsection{Couplings of the one electron states}

The modifications induced by the trapped electron-photon
interaction contain two types of processes.

 \begin{figure}
\includegraphics[width=7.7cm]{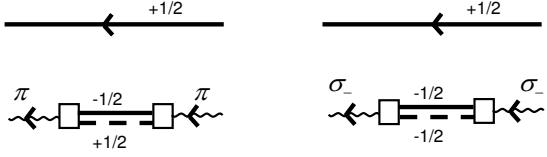}
\caption{Diagonal processes for a trapped electron with $m_s=+1/2$
spin projection along $\hat{z}$.}\label{fig5}
\end{figure}

 ({\it i}) In direct processes, the electron which recombines is
 the one of the virtual electron-hole pair coupled to the
 excitation photon. Due to Pauli exclusion, the electron of the
 virtual pair must however have a spin different from the one
 already present in the quantum dot. For an $m_s=+1/2$ electron, we
 only have the two processes shown in Figure \ref{fig5}. The energy change
 induced for these diagonal processes is
\begin{equation}
\langle\frac{1}{2}|H_{\rm eff}|\frac{1}{2}\rangle =-u+
\frac{2}{3}\frac{|\lambda_0|^2}{\delta-(u-v)}+\frac{1}{3}\frac{|\lambda_-|^2}{\delta-(u+v)}
\end{equation}

 ({\it ii}) Exchange processes are also possible between the
 trapped electron and the virtual electron-light-hole pair coupled
 to photons. Again, due to the Pauli exclusion principle, the electron spin of
 this virtual pair must be different from the trapped one. An
 electron exchange between a virtual pair coupled to a $\pi$
 photon and a $m_s=1/2$ electron leads to the diagram of Figure \ref{fig3}.a,
 while an electron exchange with a $m_s=-1/2$ electron leads to the
 diagram of Figure \ref{fig3}.b. In this exchange, the spin of the trapped
 electron flips while a $\pi$ photon is transformed into a
 circularly polarized photon.The reverse transformations being also allowed (see Figures \ref{fig3}.c and \ref{fig3}.d),
 these processes lead to the two non diagonal matrix elements
\begin{equation}
\begin{array}{lll}
\langle\frac{1}{2}|H_{\rm eff}|-\frac{1}{2}\rangle&=&
-\big(\frac{\sqrt{2}}{3}\lambda_0\lambda_{+}^*\frac{\delta+v}{(\delta+v)^2-u^2}\\
&&-\frac{\sqrt{2}}{3}\lambda_0^{*}\lambda_-\frac{\delta-v}{(\delta-v)^2-u^2} \big)\\
&=&\langle-\frac{1}{2}|H_{\rm eff}|\frac{1}{2}\rangle^*\\
\end{array}
\label{eq7}
\end{equation}
the minus sign being due to the single carrier exchange involved
in these interactions. Therefore, we obtain in
the computational basis $\{|-\frac{1}{2}\rangle$,$|\frac{1}{2}\rangle\}$
\begin{widetext}
\begin{eqnarray}
H_{\rm eff}=
\left[%
\begin{array}{cc}
 u+\frac{|\Omega_\pi|^2}{\delta-v+u}+\frac{|\Omega_+|^2}{\delta+u+v} &\frac{ \Omega_{\pi}^*\Omega_-(\delta-v)}{(\delta-v)^2-u^2}-\frac{\Omega_\pi\Omega_+^*(\delta+v)}{(\delta+v)^2-u^2} \\
\frac{\Omega_{\pi}\Omega_-^*(\delta-v)}{(\delta-v)^2-u^2}-\frac{\Omega_\pi^*\Omega_+(\delta+v)}{(\delta+v)^2-u^2}
& -u+
 \frac{|\Omega_\pi|^2}{\delta-u+v}+\frac{|\Omega_-|^2}{\delta-u-v}\\
\end{array}
\right]\label{eq9}
\end{eqnarray}
\end{widetext}
with $\Omega_{\pi}$=$\sqrt{2/3}$$\lambda_0$,
$\Omega_{+}$=$\sqrt{1/3}$$\lambda_+$ and
$\Omega_{-}$=$\sqrt{1/3}$$\lambda_-$. Please note that the time
dependance of these Rabi frequencies is implicitly considered.

Let us mention that in previous works \cite{Sham2,Sham3} a
density matrix approach has been followed to describe stimulated
Raman transitions in quantum dots. One then considers the four
following states: the two electronic levels with spin projections
in the $\hat{z}$ direction, $|+1/2\rangle$ and $|-1/2\rangle$, and
the two trion levels $|t_\uparrow\rangle$ and
$|t_\downarrow\rangle$ made of an electron pair with opposite
spins and a light hole with up and down spin projection along
$\hat{z}$ respectively. The level scheme and the different optical
transition strengths are depicted in Figure \ref{fig2}.b. Please
note that to preserve proper anti-commutation relations between
fermionic operators, the optical transitions have in this point of
view different signs compared to the previous expressions. A model
Hamiltonian can thus be built in the $\{|1/2\rangle$,
$|-1/2\rangle$, $|t_\downarrow\rangle$, $|t_\uparrow\rangle\}$
basis, and turning into the rotating frame, the equations of
motion in the restricted $\{|+1/2\rangle$, $|-1/2\rangle\}$
subspace are obtained after adiabatic elimination of the virtually
populated trion levels. This procedure provides the proper
effective Hamiltonian in second order perturbation theory, but let us
stress that care must be taken in order to correctly account for
non degeneracy of the ground states.

\section{Building single qubit rotations}
Consider the geometry in Fig. \ref{fig1} with the $\hat{z}$ axis defined as the axis of symmetry of the quantum dot.
 The addressing laser propagates with a wavevector ${\bf k}$ at an angle $\beta$ with respect to the $\hat{z}$ axis in the $(Oxz)$ plane.
 We choose an elliptical polarization of the electric field with respect to a primed coordinate frame $(\hat{z'}\equiv\hat{k})$ given by
 ${\bf E}'({\bf x},t)=E e^{i({\bf k}\cdot {\bf x}-\omega t)}(\cos{\gamma}{\bf e}'_-+e^{i\phi}\sin{\gamma}{\bf e}'_+)$.
The field components in the spherical basis of the unprimed coordinate
system of the dot are given by ${\bf
E}({\bf x},t)=D^{(1/2)\dagger}(\beta){\bf E'}({\bf x},t)$, where $D^{(1/2)}$ is the
reduced Wigner rotation matrix for a spin-1/2 representation of
$SU(2)$ \cite{Sakuri}.

\begin{figure}
\includegraphics[width=\columnwidth]{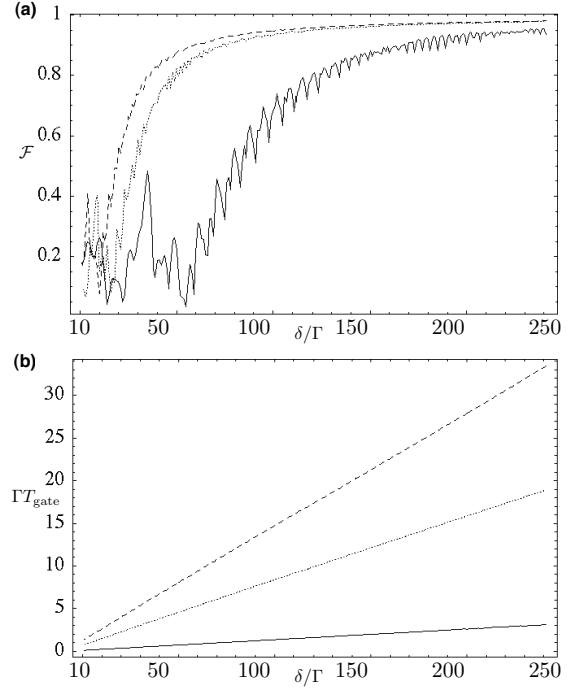}
\caption{\label{fig6}Single qubit gate performance as a function of field detuning
$\delta$.  The system parameters are
$u/\hbar\Gamma=0.1,v/\hbar\Gamma=0.2,\beta=\pi/10$ and three different field strengths are plotted:  solid lines correspond to
 $\Omega/\Gamma=50$,
dotted lines $\Omega/\Gamma=20$, and dashed lines $\Omega/\Gamma=15$.
(a)  Worst case fidelity for
a single qubit operation.  (b)
Maximum gate time.}
\end{figure}

During single qubit gate operations, there will be decoherence
induced by off resonant laser coupling to unstable trion states.
We describe such decay processes by a phenomenological parameter
$\Gamma$. The effect of this decay on gate performance is captured
by constructing an effective non Hermitian Hamiltonian
$\tilde{H}_{\rm eff}$ from $H_{\rm eff}$ by the replacement
$\delta\rightarrow \delta-i\Gamma/2$.  The dot-photon interaction
can then be represented as a vector in the operator space spanned
by the Pauli operators. Ignoring the component of the identity
operator, we write
$\tilde{H}_{\rm eff}=\hbar\Gamma
(\vec{h^R}+i\vec{h^I})\cdot\vec{\sigma}$.  In the limit $u,v\ll
\delta$, the real parts of the dimensionless coupling vectors are
\begin{equation}
\begin{array}{lll}
h_x^R&=&\frac{\kappa_3\kappa_4^2 \sin{\beta}\cos{\beta}(4\cos{\phi}\cos{\gamma}\sin{\gamma}-2)}{4\kappa_3^2+1}\\
h_y^R&=&\frac{2\kappa_3\kappa_4^2\sin{\beta}\sin{2\gamma}\sin{\phi}}{4\kappa_3^2+1}\\
h_z^R&=&\kappa_1-\frac{\kappa_3\kappa_4^2\cos{\beta}\cos{2\gamma}}{4\kappa_3^2+1}
\end{array}
\end{equation}
The system parameters, which are assumed fixed during gate operations, are $\kappa_1=u/\hbar\Gamma, \kappa_2=v/\hbar\Gamma,\kappa_3=\delta/\Gamma,\kappa_4=\Omega/\Gamma$.  The Rabi frequency $\Omega=-\sqrt{2/3}PEe/m\omega$ and $\Gamma$ is the parameter describing decay of the electron-light-hole pair.

Single qubit gates can be performed by using a sequence of laser pulses with varying polarization parameters $(\gamma,\phi)$.  Because these pulses share the same wavevector ${\bf k}$, they can be obtained from the same source and used in the same geometric configuration relative to the dot.
The nature of the sequence follows from the canonical decomposition of any unitary $U \in SU(2)$:
\begin{equation}
U=e^{i\xi \hat{n}\cdot\vec{\sigma}}=(e^{-i\mu_3 \hat{n}_1\cdot\vec{\sigma}})(e^{-i\mu_2 \hat{n}_2\cdot\vec{\sigma}})(e^{-i\mu_1 \hat{n}_1\cdot\vec{\sigma}}),
\label{Euler1}
\end{equation}
provided the Bloch vectors are orthogonal: $\hat{n}_1\cdot\hat{n}_2=0$.  Expanding this generalized Euler decomposition gives the following relations
\[
\begin{array}{lll}
\cos{\xi}&=&\cos{\mu_2}\cos(\mu_1+\mu_3)\\
\hat{n}\sin{\xi}&=&-\hat{n}_1\cos{\mu_2}\sin(\mu_1+\mu_3)-\hat{n}_2\sin{\mu_2}\cos(\mu_1-\mu_3)\\
&&+\hat{n}_1\times \hat{n}_2\sin{\mu_2}\sin(\mu_1-\mu_3)
\end{array}
\label{Euler2}
\]
These equations can be inverted and without loss of generality we can choose $\mu_i>0$.  Hence it is only required that for a fixed set of parameters $\{\kappa_i\}$, we find a pair of non zero vectors satisfying $\vec{h^R}(\gamma_1,\phi_1)\cdot\vec{h^R}(\gamma_2,\phi_2)=0$.

The process fidelity can be quantified by the overlap of the
target unitary $U$ with the implemented operator. We model the
implemented operator as the non unitary evolution generated by
three sequential evolutions by the non-Hermitian Hamiltonian
$\tilde{H}_m$ acting on the joint electron-trion space:
\begin{equation}
\begin{array}{lll}
\tilde{H}_m&=&\hbar\Gamma\Big[\kappa_1 (\ket{-\frac{1}{2}}\bra{-\frac{1}{2}}-\ket{\frac{1}{2}}\bra{\frac{1}{2}})\\
&+&(\kappa_2-\kappa_3-i/2)
\ket{t_{\downarrow}}\bra{t_{\downarrow}}-(\kappa_2+\kappa_3+i/2)
\ket{t_{\uparrow}}\bra{t_{\uparrow}}\\
&+&\Big(\frac{\kappa_4}{\sqrt{2}}(e^{-i\phi}\sin{\gamma}\cos^2{\frac{\beta}{2}}+\cos{\gamma}\sin^2{\frac{\beta}{2}})\ket{-\frac{1}{2}}\bra{t_{_\uparrow}}

\\
&+&\frac{\kappa_4}{\sqrt{2}} \sin{\beta}(e^{-i\phi}\sin{\gamma}-\cos{\gamma})
(\ket{-\frac{1}{2}}\bra{t_{_\downarrow}}-\ket{\frac{1}{2}}\bra{t_{_\uparrow}})\\
&+&\frac{\kappa_4}{\sqrt{2}}(\cos{\gamma}\cos^2{\frac{\beta}{2}}+e^{-i\phi}\sin{\gamma}\sin^2{\frac{\beta}{2}})\ket{\frac{1}{2}}\bra{t_{_\downarrow}}\Big)+h.c.\Big]
\end{array}
\end{equation}
Hence we adopt the following measure of operator fidelity:
\begin{equation}
\begin{array}{lll}
\mathcal{F}&=&\frac{1}{2}\Big|\mathrm{Tr}\Big[U^{\dagger}P_g e^{-it_3\tilde{H}_m/\hbar}e^{-it_2\tilde{H}_m/\hbar}e^{-it_1\tilde{H}_m/\hbar}P_g\Big]\Big|
\end{array}
\end{equation}
where
\[
\begin{array}{lll}
U&=&\exp\big[-it_3\vec{h^R}(\gamma_1,\phi_1)\cdot\vec{\sigma}\big]\exp\big[-it_2\vec{h^R}(\gamma_2,\phi_2)\cdot\vec{\sigma}\big]\\
&&\times\exp\big[-it_1\vec{h^R}(\gamma_1,\phi_1)\cdot\vec{\sigma}\big].
\end{array}
\]
and $P_g$ is the projector onto the computational basis.  In terms of the parameterization for $U$, $t_1=\mu_1/\Gamma|\vec{h^R}(\gamma_1,\phi_1)|$,
$t_2=\mu_2/\Gamma|\vec{h^R}(\gamma_2,\phi_2)|$, and $t_3=\mu_3/\Gamma|\vec{h^R}(\gamma_1,\phi_1)|$.
The worst case fidelity for building a generic gate is estimated by assuming $\mu_i=\pi$ $\forall i$ (see Fig. \ref{fig6}).

In order to construct arbitrary single qubit gates using Eq. \ref{Euler1} we can set $\phi_1=\pi,\phi_2=0$ such that $h_y(\gamma_1,\phi_1)=h_y(\gamma_2,\phi_2)=0$.  Furthermore, it suffices to choose $\gamma_1=\gamma_2+\pi/2\equiv\gamma$.   For a given set $\{\kappa_i\}$ it is then possible to solve for $\gamma$ such that $\vec{h^R}(\gamma,\pi)\cdot\vec{h^R}(\gamma-\pi/2,0)=0$.  For example consider a setup with $\kappa_1=0.1,\kappa_2=0.2,\kappa_3=250,\kappa_4=20,\beta=\pi/10$.  The following sequence simulates the Hadamard gate $H$ (up to a global phase):
\begin{equation}
\begin{array}{lll}
H&=&\frac{1}{\sqrt{2}}\left(\begin{array}{cc}1 & 1 \\1 & -1\end{array}\right)\\
&=&ie^{-i\Gamma t_3 \vec{h^R}(\gamma,\pi)\cdot\vec{\sigma}}e^{-i\Gamma t_2 \vec{h^R}(\gamma-\pi/2,0)\cdot\vec{\sigma}}e^{-i\Gamma t_1 \vec{h^R}(\gamma,\pi)\cdot\vec{\sigma}},
\end{array}
 \end{equation}
in the basis $\{\ket{-1/2},\ket{1/2}\}$.  Here the field setting is $\gamma=0.4324\pi$ and the laser pulse times are:  $\Gamma t_3=4.1470, \Gamma t_2=1.4122,\Gamma t_1=4.1470$.  The gate fidelity is $\mathcal{F}=0.9891$.  Another gate, the phase gate $P=e^{-i\frac{\pi}{8}\sigma^z}$, can likewise be simulated using the same system parameters but with the pulse times:  $\Gamma t_3=3.1130, \Gamma t_2=0.5566,\Gamma t_1=3.1130$.  The gate fidelity is
$\mathcal{F}=0.9902$.  The group generated under multiplication of the set $S=\{H,P\}$ is dense in $SU(2)$ and $S$ complemented by the two qubit $\mathtt{CNOT}$ gate is sufficient for universal quantum computation \cite{Boykin}.

Faster gate times are possible using larger field strengths and smaller detunings.
For instance, for the same system parameters but with
$\kappa_3=200,\kappa_4=50$ we simulate the Hadamard with fidelity $\mathcal{F}=0.9709$ using the field setting $\gamma=0.4263\pi$ and laser pulse times:  $\Gamma t_3=0.2052, \Gamma t_2=0.4303,\Gamma t_1=0.2052$.  For a light hole lifetime of $50{\rm ps}$ this corresponds to a gate time
$T_{\rm gate}\approx 42{\rm ps}$ using fields detuned by $16.7 {\rm meV}$ from optical resonance.  Fidelities and gate times for generic single qubit unitaries are plotted in Fig. \ref{fig6}.


\section{Conclusions}

We have proposed a simple way to experimentally implement
arbitrary rotations with a single electron trapped in a quantum
dot. As in many other works, the qubit of information is encoded
in the electron spin. However, unlike in previous proposals, our
procedure is based on a single laser frequency, in a one axis
geometry. The latter is the laser wave vector which is set tilted
with respect to the quantum dot growth direction. When the laser
is detuned from the quantum dot electron-light-hole optical
transition, a coherent coupling between the two spin states of the
trapped electron can be induced. Moreover, in the particular
experimental situation where a weak magnetic field is applied
along the growth axis, setting appropriate elliptical
polarizations for the addressing laser allows one to perform arbitrary
single qubit rotations. These reach $>97\%$ fidelities, in $50{\rm
ps}$, even in the worst scenario for which light-holes in quantum
dots would exhibit a lifetime as short as in quantum wells.
Furthermore, unitary operations fidelities are robust against the
exact Zeeman shift of electron and hole levels, the laser detuning
being in any case very large compared to these parameters.
According to our analysis, in quantum dots, light-holes could
become a favorable candidate for optical implementation of quantum
computation protocols.

We have not addressed the issue of state preparation here.  One possibility
 is to tune a field near resonant with the light-hole states and monitor
 spontaneous emission events.  A null result projects the mixed qubit
 state into the non degenerate null space of the Hamiltonian $\tilde{H}_m$
 spanned by the computation basis states.  The requirements for robust
 state preparation would be a decay rate fast compared to coherent dynamics in
 the $4$ level system as well as high detector efficiency.

\begin{acknowledgments}
F.D. appreciates many stimulating discussions with M. Combescot.
F.D. and GKB received support from the Austrian Science 
Fund (FWF). 
\end{acknowledgments}


\begin{thebibliography}{99}
\bibitem{note} Trion states in principle refer to a bound exciton
interacting with an electron. However, in a quantum dot the term
trion is used to describe two electrons and a hole bound by
confinment.
\end{thebibliography}
\end{document}